\documentclass[12pt]{article}
\usepackage{epsfig}

\def\beq{\begin{equation}}
\def\eeq#1{\label{#1}\end{equation}}
\def\eeqn{\end{equation}}

\def\beqa{\begin{eqnarray}}
\def\eeqa#1{\label{#1}\end{eqnarray}}
\def\eeqan{\end{eqnarray}}

\let\bar=\overbar

\def\Dslash{\not{\hbox{\kern-4pt $D$}}}
\def\dslash{\not{\hbox{\kern-2pt $\del$}}}

\def\msb{{\bar{\ssstyle M \kern -1pt S}}}

\def\Title#1{\begin{center} {\Large {\bf #1} } \end{center}}

\begin{document}

\Title{Lepton-Flavor Violation and Physics beyond the Standard Model}

\bigskip\bigskip


\begin{raggedright}  

{\it Junji Hisano\index{HISANO,J}\\
Department of Physics\\
Nagoya University \\
Nagoya 464-8602, Japan}
\bigskip\bigskip
\end{raggedright}

\section{Introduction}
\label{}

Nowadays various charged lepton-flavor violating (LFV) processes are
being searched for in experiments. Charged LFV searches have been
studied since muons were discovered. Tau LFV searches were also
started soon after tau leptons were discovered. However, charged LFV
processes have not yet been discovered, and the upper bounds on the
branching ratios of the processes are being updated.

We know that the lepton-flavor symmetries are not exact in nature, since
the neutrino oscillation was discovered. The neutrino oscillation is
induced by the finite but tiny neutrino masses. On the other hand,
the charged LFV processes derived due to the neutrino masses have
negligible event rates. This comes from the GIM mechanism in leptonic
sector. In fact, $Br(\mu\rightarrow e \gamma)$ is limited to be below
$10^{-54}$ in the standard model (SM)  with the tiny neutrino masses.

On the other hand, it is considered that the standard model should be
a low-energy effective theory and new physics may appear at TeV
scale. Now we know that the lepton flavor symmetries are not exact in
nature, and we guess that the symmetries may be broken in the
model. In that case, the charged LFV processes are predicted with
branching ratios accessible to experiments in near future.

This year a Higgs-like particle $h$ has been discovered at the LHC with
mass around 125~GeV. On the other hand, the models
beyond the SM are severely constrained from null results for searches
for exotic events, such as events with missing $E_T$, at the LHC. The
supersymmetric standard model (SUSY SM) is the leading candidate for
physics beyond the SM at TeV scale, while it has been found at the LHC
experiments that the squark and gluino masses are bounded above about
1~TeV. This is consistent with the observation of the Higgs-like
particle, since the large radiative correction to the Higgs boson mass
is needed in the model in order for the Higgs boson to have mass
around 125~GeV.

There may still be some clues for physics beyond the standard model at
TeV scale in the non-colored sector. Mild excess of the branching
ratio of $h\rightarrow \gamma \gamma$ may imply existence of a light
non-colored particle \cite{Carena:2012gp}, since the branching ratios
to the weak bosons are consistent with the SM prediction.  Stau in the
SUSY SM is a possible candidate for the non-colored particle coupled with the
Higgs boson. The large correction to the branching ratio requires the
stau mass close to the LEP bound in the SUSY SM while such parameter
region should be needed to be studied more. The muon $(g-2)$ has a
long history after the deviation from the SM prediction was reported. It also prefers the light non-colored
particles, such as slepton and chargino/neutralinos.

Non-colored sector may include new physics related to those anomalies,
while null results for new physics searches at the LHC might imply
that physics beyond the standard model is much higher than TeV scale.
In both cases the charged LFV searches would be important to search
for physics beyond the SM.

Now I would like to discuss about two topics related to the charged
LFV processes. First is whether the discovered Higgs-like particle is
really the Higgs boson in the minimal SM.  Next is whether models beyond the
SM still have observable prediction for the charged LFV processes in
future experiments. 

\section{Non-minimal Higgs boson}

\begin{figure}[t]
  \begin{center}
    \epsfig{file=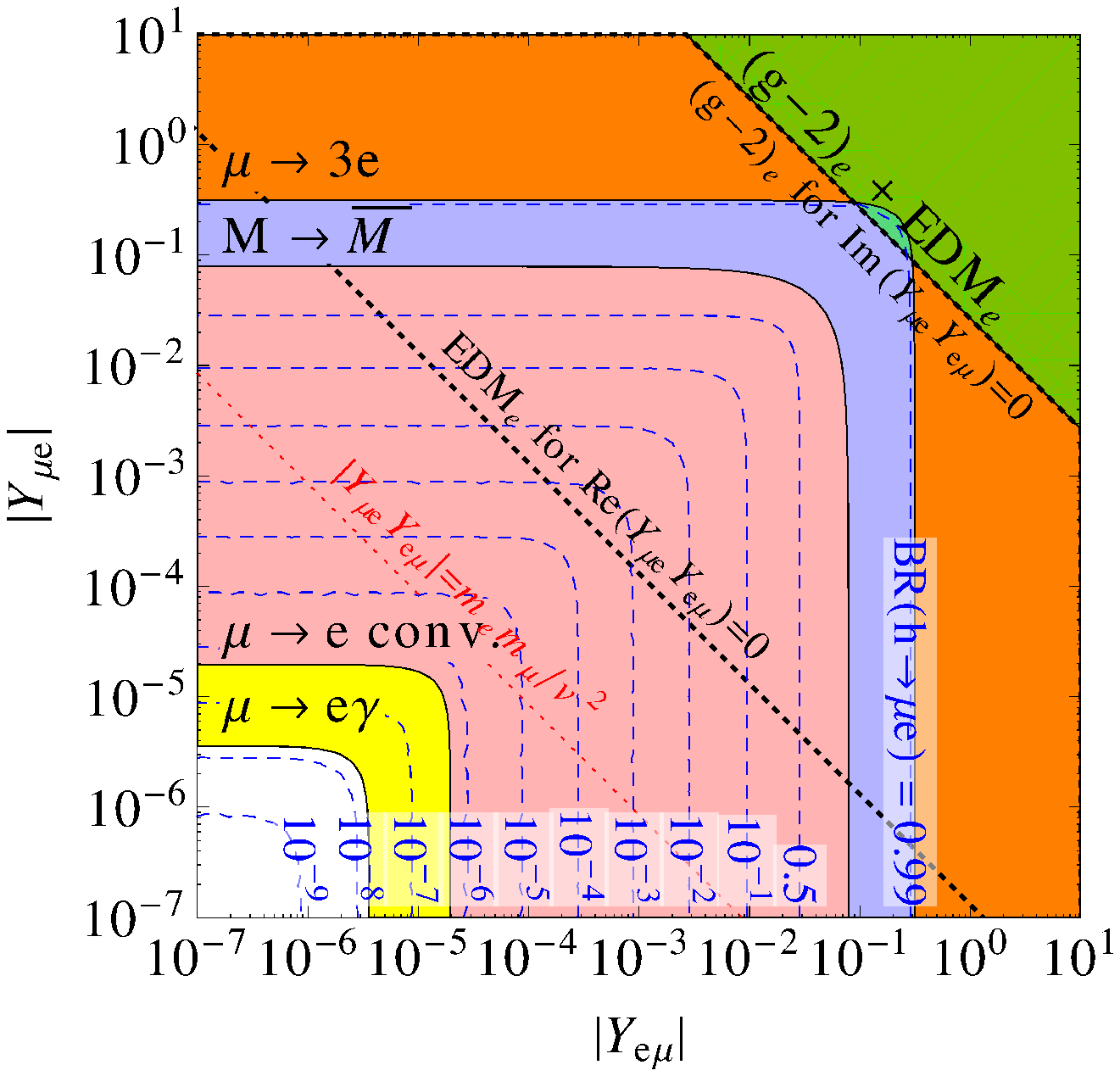,height=1.5in}
    \epsfig{file=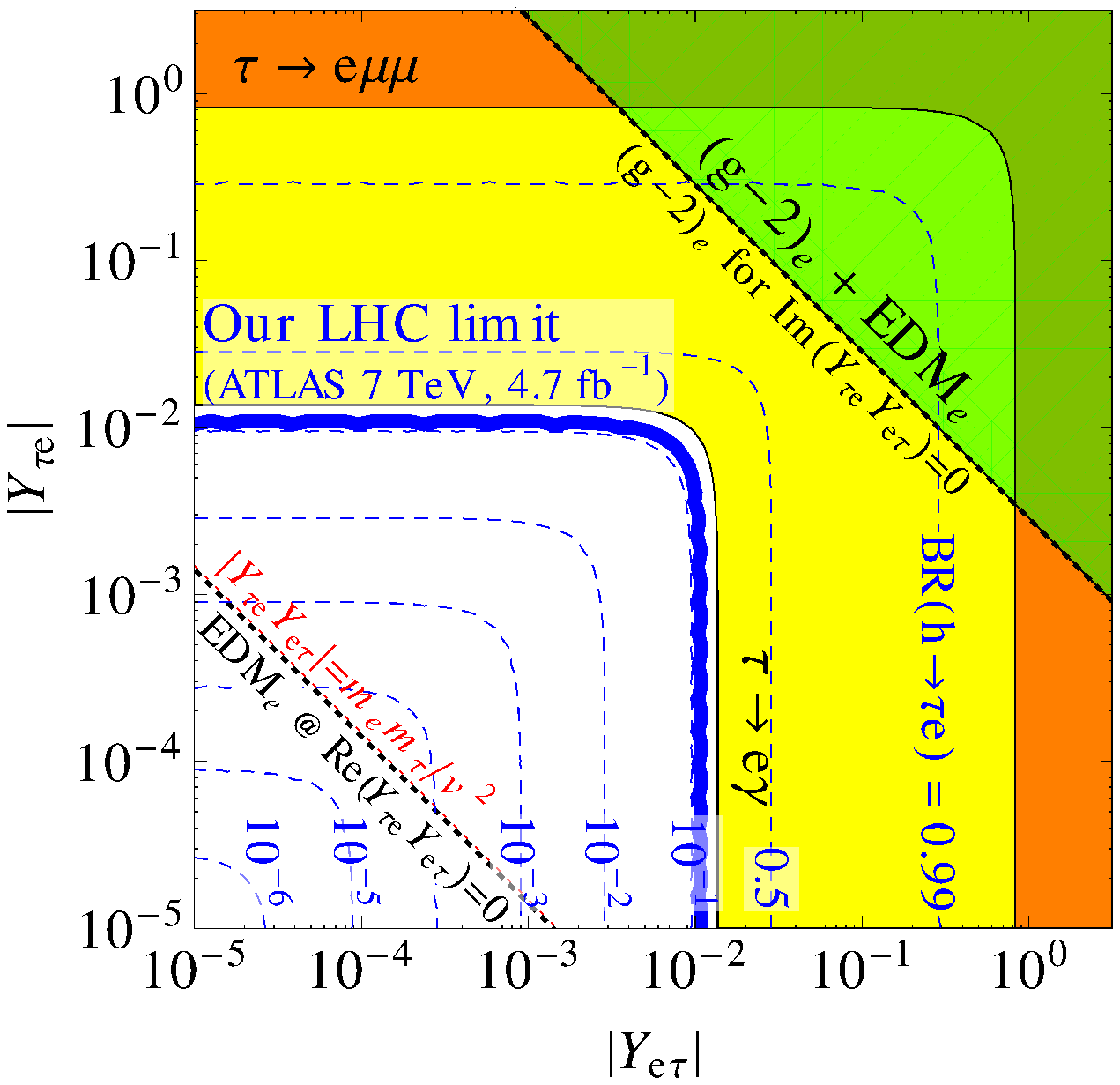,height=1.5in}
    \epsfig{file=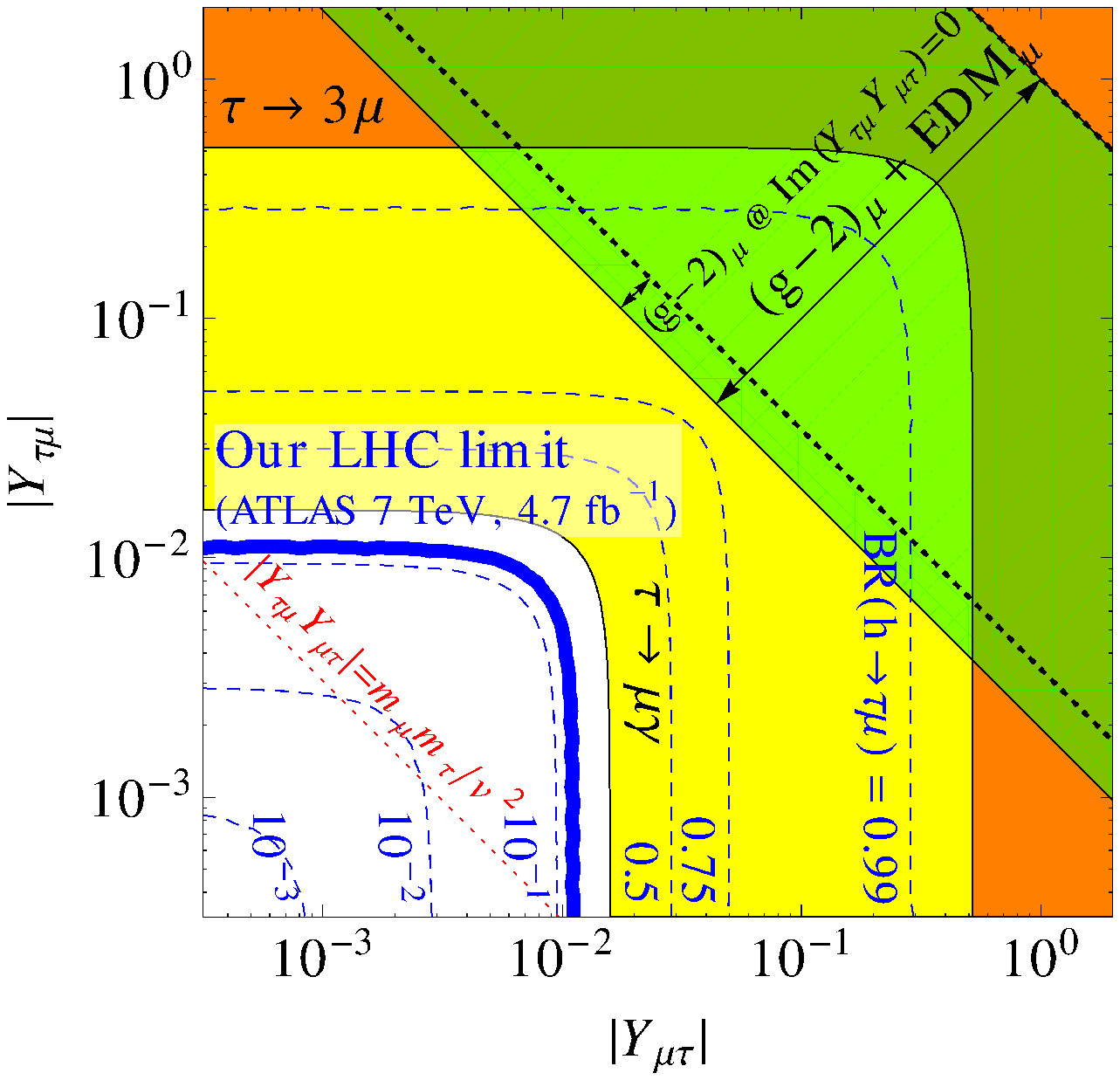,height=1.5in}
  \end{center}
  \caption{Constraints on LFV Yukawa coupling constants of the Higgs boson.
Figures come from Ref.~\cite{Harnik:2012pb}.}
  \label{fig:Higgs}
\end{figure}

Now let us consider possibilities that the observed Higgs-like
particle is a non-minimal Higgs boson. Even in the standard model we
may introduce LFV Yukawa coupling of the Higgs
boson, if the higher dimensional interactions are introduced.  The
LFV in the Yukawa coupling is studied well, and it
is found that flavor violation between first and second
generations is severely constrained by $\mu\rightarrow e \gamma$ and
also the $\mu$--$e$ conversion in nuclei. On the other hand, flavor
violations between $\tau$ and $\mu$ and between $\tau$ and $e$ are
still much less constrained. Thus, we may observe $h\rightarrow \tau \mu$
or $h\rightarrow \tau e$ at the LHC \cite{Blankenburg:2012ex,Harnik:2012pb}.

In Ref.~\cite{Harnik:2012pb}, Harnik, Kopp and Zupan evaluated bounds on the tau LFV decays of the
Higgs boson at the LHC. They used an analysis of $h\rightarrow
\tau \tau$ using leptonic tau decay by the Atlas group. They argued
that the branching ratio of  $h\rightarrow \tau \mu$ is already
constrained to be smaller than about 0.1 at the Atlas experiment.

This gives a constraint on the LFV Yukawa coupling, comparable to
those from low-energy experiments.  In Fig.~\ref{fig:Higgs},
constraints on LFV Yukawa coupling constants from low-energy experiments and
also branching ratio of the LFV Higgs decay are shown.  The LFV Yukawa
coupling constants are given by
\begin{eqnarray}
-{\cal L}_{\rm LFV}
&=&
 \sum_{i,j=e,\mu,\tau}  m_i \bar{f}^i_L {f}^i_R+Y_{ij} \bar{f}^i_L {f}^j_R h + {\rm h.c.}.
\end{eqnarray}
Thus the bounds on $Y_{\tau\mu}$ and $Y_{\mu\tau}$ from the LHC and
low-energy experiments, such as $\tau\rightarrow 3\mu$, are
comparable. For the $\tau$--$e$ case, the constraints from the LHC and
low-energy experiments are also comparable. However, region above the
diagonal red lines might be disfavored from naturalness in the Yukawa
coupling.  For the $\mu$--$e$ case, the constraints from low-energy
experiments are too strong, then $h\rightarrow \mu e$ cannot be
discovered at the LHC. It is quite interesting that the LHC or the B
factories are competitive in studies of the LFV Yukawa coupling of the
Higgs boson.

\section{SUSY seesaw model} 

\begin{figure}[t]
  \begin{center}
    \epsfig{file=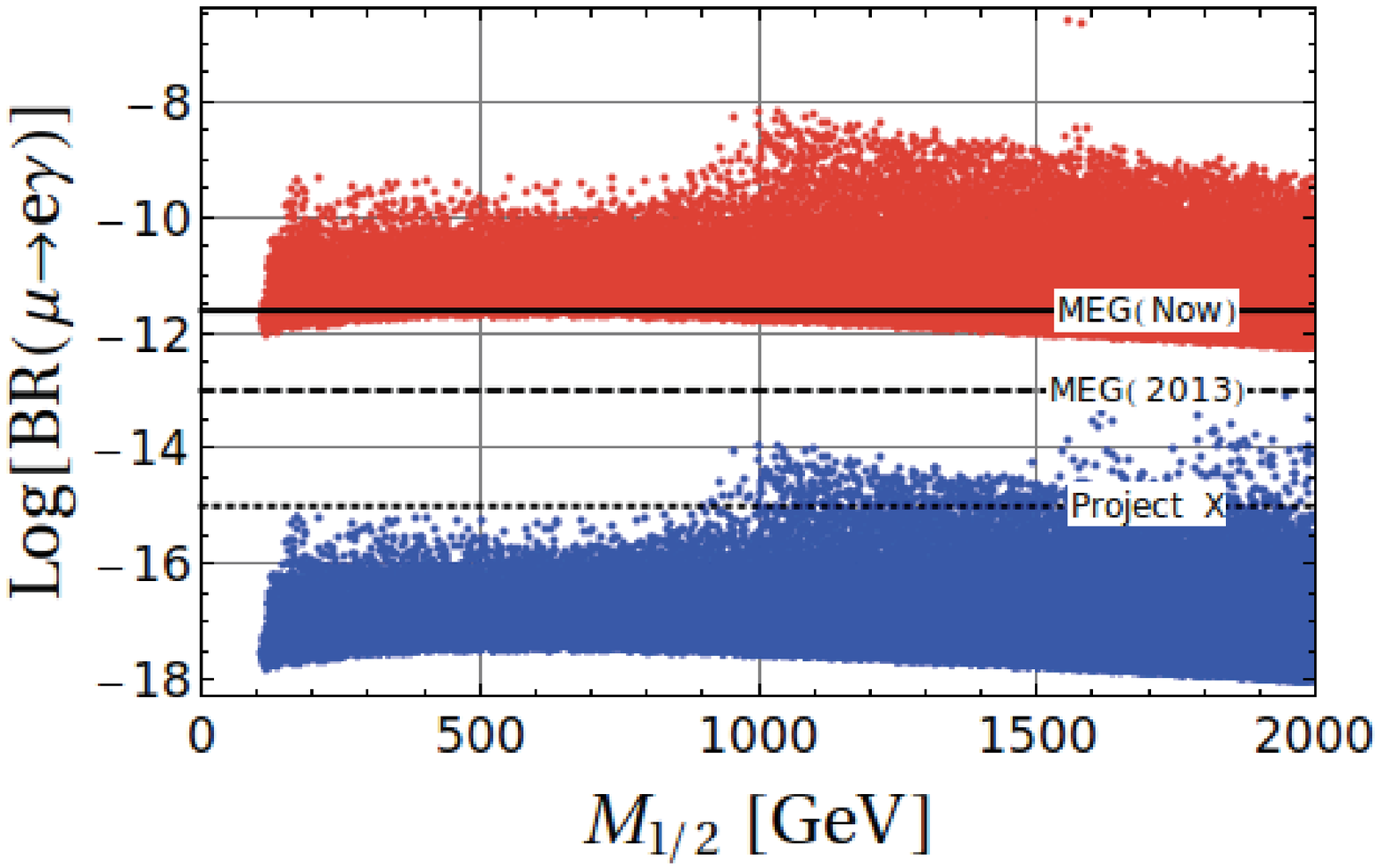,height=1.5in}
    \epsfig{file=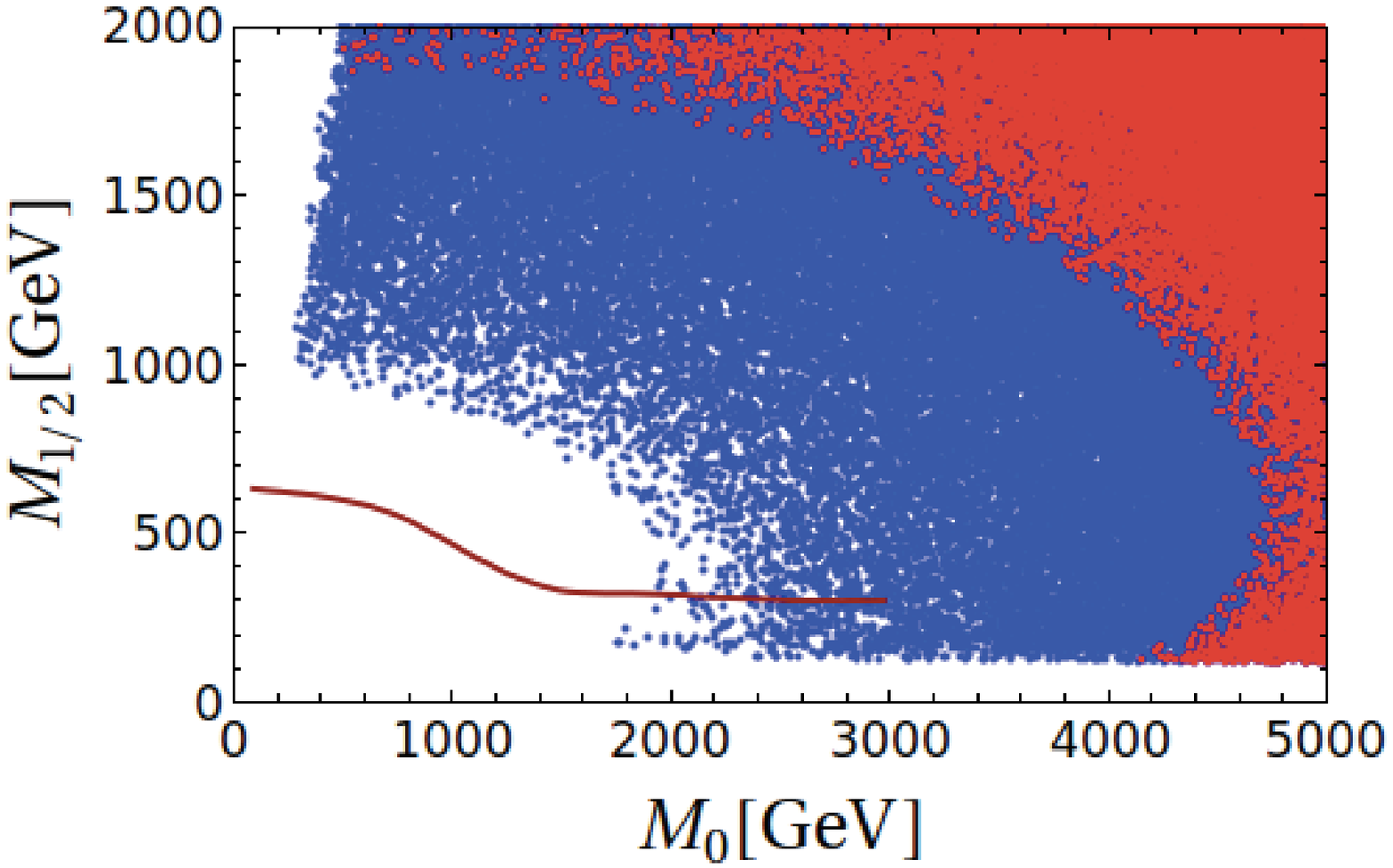,height=1.5in}
  \end{center}
  \caption{(Left) $Br(\mu\rightarrow e \gamma)$ in SUSY seesaw
    model. Red and blue are for the PMNS and CKM cases,
    respectively. (Right) Allowed region from $\mu\rightarrow e
    \gamma$ in the PMNS and CKM cases. These figures are from
    Ref.~\cite{Calibbi:2012gr}.}
  \label{fig:seesaw1}
\end{figure}

Next we move to models beyond the standard model. The leading candidate
for physics beyond the SM is the SUSY SM. In the SUSY SM, the new
flavor violation is introduced in slepton and squark mass matrices as
the SUSY breaking terms. When the sleptons have off-diagonal terms in
the mass matrices, the charged LFV processes are predicted.  Here, we
discuss the charged LFV processes in the SUSY SM, while many (non-SUSY) models
beyond the SM also predict the charged LFV processes and some models
are severely constrained already by the experimental bounds, such as 
studies in Refs.~\cite{Agashe:2006iy,Blanke:2007db}.

The constraints on the SUSY models from the hadronic and leptonic
flavor-changing processes are quite severe. This is called the SUSY
flavor problem. The representative proposal to solve the problem is
the universal scalar mass hypothesis.  The hypothesis is realized when
supersymmetry is spontaneous broken in the hidden sector and the SUSY
breaking is transmitted to the SUSY SM by some flavor-independent
messenger interaction, such as gravity or gauge interaction. The
off-diagonal terms in the squark and slepton mass matrices are
suppressed automatically.

Even in this setup, the charged LFV processes may be predicted. If
some LFV interaction is active below the messenger scale, the LFV mass
terms for sleptons are radiatively generated. SUSY GUTs \cite{Barbieri:1994pv} and SUSY
seesaw models \cite{Borzumati:1986qx} introduce the LFV interactions. Then, if the SUSY
breaking is transmitted by gravitational interaction, charged LFV processes
may be good prediction.

Now we consider the SUSY seesaw model, in which right-handed neutrinos
are introduced in order to predict tiny neutrino masses. In this
model, the neutrinos have the LFV Yukawa coupling. The charged LFV
processes depend on the structure of the neutrino Yukawa coupling. The
Masiero's group updated their result including the LHC constraints, and
observed neutrino mixing angle $U_{e3}$ \cite{Calibbi:2012gr}. They
studied two cases, the first one is that neutrino Yukawa is given by the
CKM matrix and diagonal term of up-quark Yukawa coupling at the GUT
scale. In the second one they use the PMNS matrix instead of the CKM one. For the SUSY
breaking parameters they adopt the mSUGRA, which is a working
hypothesis based on the gravity mediation. Left figure of
Fig.~\ref{fig:seesaw1} shows that the PMNS case predicts larger
$Br(\mu\rightarrow e \gamma)$ than the CKM one. Almost all parameter regions they studied
are excluded by the current bound derived by the MEG experiment if the
PMNS matrix is assumed in the neutrino Yukawa coupling. In right
figure of Fig.~\ref{fig:seesaw1}, region allowed by $\mu\rightarrow e
\gamma$ for PMNS and CKM cases are shown in the plane of
$M_0$ and $M_{1/2}$, which are input parameters of the mSUGRA. The region
above red line is allowed by the direct SUSY particle searches. The
Higgs boson mass are chosen to be consistent with the discovered
Higgs-like boson. It is found that the $\mu\rightarrow e \gamma$
constraint may still be stronger than those from the LHC in this setup.
Notice that smaller $Br(\mu\rightarrow e \gamma)$ is predicted in the
non-universal Higgs mass scenario which is a variant of mSUGRA, since it has
more input parameters.

\begin{figure}[t]
  \begin{center}
    \epsfig{file=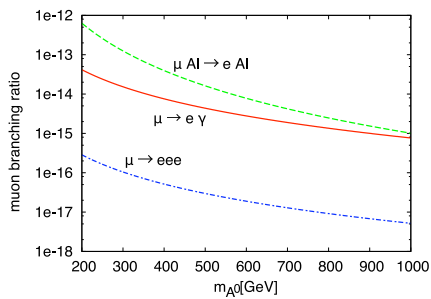,height=1.5in}
  \end{center}
  \caption{Branching ratios for muon LFV processes are shown assuming the Higgs
boson contribution is dominant in the SUSY SM. Figure comes from Ref.~\cite{Hisano:2010es}.}
  \label{fig:seesaw3}
\end{figure}

If SUSY particle masses are larger than $O(1-10)$~TeV, SUSY contribution
to flavor changing processes may be much suppressed even if squark and
slepton mixing are not suppressed at all. Notice that, even in that
case, the Higgs-boson exchange contributes to the charged LFV processes, since SUSY SM
has two doublet Higgs bosons \cite{Babu:2002et}. The LFV Yukawa coupling of the Higgs bosons are generated after
integrating out SUSY particles at one-loop level. In Fig.~\ref{fig:seesaw3}
branching ratios for muon LFV processes are shown assuming the Higgs
boson contribution is dominant. As the result, the $\mu$--$e$ conversion
in nuclei is the largest among the processes.  

Here, we get a lesson: When a new particle is found, we need to check
whether it has LFV interaction or not.

\section{Summary}

Searches for the charged LFV process have a long history, while their
importance is increasing now. Notice that the studies should be
important whether new physics is discovered or not at the LHC, since
they have potential to probe physics higher than TeV scale.




\nocite{*}
\bibliographystyle{elsarticle-num}
\bibliography{martin}


\end{document}